%% file: main.tex
\documentclass[conference]{IEEEtran}
\IEEEoverridecommandlockouts
\usepackage{cite}
\usepackage{listings}
\usepackage{comment}
\usepackage[dvipsnames]{xcolor}
\usepackage{subfigure}
\definecolor{codegreen}{rgb}{0,0.6,0}
\definecolor{codegray}{rgb}{0.5,0.5,0.5}
\definecolor{codepurple}{rgb}{0.58,0,0.82}
\definecolor{backcolour}{rgb}{0.95,0.95,0.92}
\usepackage{listings}
\usepackage{courier}

\usepackage{pifont}

\lstdefinestyle{mystyle}{
    backgroundcolor=\color{backcolour},   
    commentstyle=\color{Red},
    keywordstyle=\color{Blue},
    numberstyle=\tiny\color{codegray},
    stringstyle=\color{codepurple},
    basicstyle=\ttfamily\scriptsize,
    breakatwhitespace=false,         
    breaklines=true,  
    captionpos=b,                    
    keepspaces=true,                 
    numbers=none,                    
    numbersep=5pt,                  
    showspaces=false,                
    showstringspaces=false,
    showtabs=false,  
    tabsize=2
}

\usepackage{tikz}
\usetikzlibrary{tikzmark}

\usepackage{amsmath,amssymb,amsfonts}
\usepackage{algorithmic}
\usepackage{graphicx}
\usepackage{textcomp}
\usepackage{xcolor}
\usepackage{comment}
\usepackage{multirow}
\newcommand{\sentaur}{{\sffamily\small SENTAUR}}
\def\BibTeX{{\rm B\kern-.05em{\sc i\kern-.025em b}\kern-.08em
    T\kern-.1667em\lower.7ex\hbox{E}\kern-.125emX}}
\begin{document}

\title{SENTAUR: \underline{S}ecurity \underline{E}nha\underline{N}ced \underline{T}rojan \underline{A}ssessment Using LLMs Against \underline{U}ndesirable \underline{R}evisions}

\author{%
\IEEEauthorblockN{Jitendra Bhandari, Rajat Sadhukhan, Prashanth Krishnamurthy, Farshad Khorrami, Ramesh Karri}
\IEEEauthorblockA{
\textit{New York University, New York, USA}
}
}

\maketitle

\begin{abstract}
A globally distributed  IC supply chain brings risks due to untrusted third parties. The risks span inadvertent use of hardware Trojan (HT), inserted Intellectual Property (3P-IP) or Electronic Design Automation (EDA) flows. HT can introduce stealthy HT behavior, prevent an IC work as intended, or leak sensitive data via side channels. To counter HTs, rapidly examining  HT scenarios is a key requirement. While Trust-Hub benchmarks are a good starting point to assess defenses, they encompass a small subset of manually created HTs within the expanse of HT designs. Further, the HTs may disappear during synthesis. We propose a large language model (LLM) framework \sentaur{} to generate a suite of legitimate HTs for a Register Transfer Level (RTL) design by learning its specifications,  descriptions, and natural language descriptions of HT effects. Existing tools and benchmarks are limited; they need a learning period to construct an ML model to mimic the threat model and are difficult to reproduce.  \sentaur{} can swiftly produce HT instances by leveraging LLMs without any learning period and sanitizing the HTs facilitating their rapid assessment. Evaluation of \sentaur{} involved generating effective,  synthesizable, and practical HTs from TrustHub and elsewhere, investigating impacts of payloads/triggers at the RTL. While our evaluation focused on HT insertion, \sentaur{} can generalize to automatically transform an RTL code to have defined functional modifications.

\end{abstract}

\begin{IEEEkeywords}
Hardware Trojan Detection, Hardware Security, Golden Reference-Free, Large Language Model
\end{IEEEkeywords}

\input{files/intro}


\input{files/prompt}
\input{files/results}

\bibliographystyle{IEEEtran.bst}      
\bibliography{bibliography.bib}

\end{document}

%% file: files/intro.tex
\section{Introduction}
Chip manufacturers face increasing challenges due to the scale and complexity of System-on-Chip (SoC) designs specifically targeted for modern embedded systems and Internet-of-Things (IoT) devices. Consequently, SoC designers, under time-to-market pressures and resource limitations, have turned to outsourcing hardware designs and utilizing Third-Party Electronic Design Automation tools or Intellectual Property cores (combined 3P-EDA/IP) from global vendors. 
This outsourcing presents numerous benefits, including cost reduction by minimizing internal team overheads, utilization of specialized skills through specialist firms for efficient resource allocation, faster development for crucial time-to-market scenarios, and access to a wider global talent pool with diverse skill sets.

\begin{figure*}[!ht]
\centering
\subfigure[Scope of \sentaur{} Flow in IC Supply Chain for HT Analysis]{\includegraphics[scale=0.4]{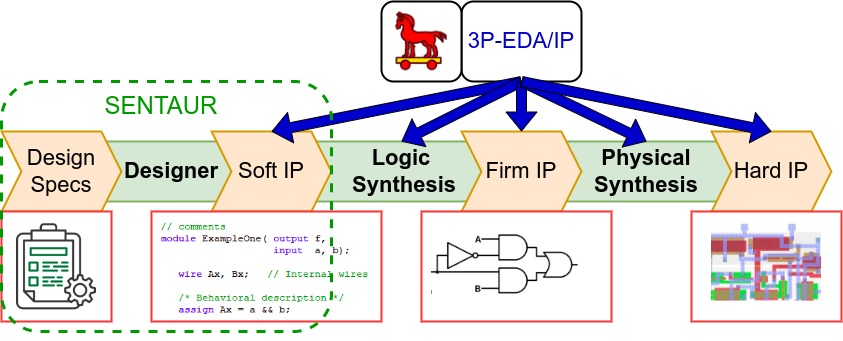}}
\hspace*{-0.15in}
\subfigure[\sentaur{} Flow]{\includegraphics[scale=0.25]{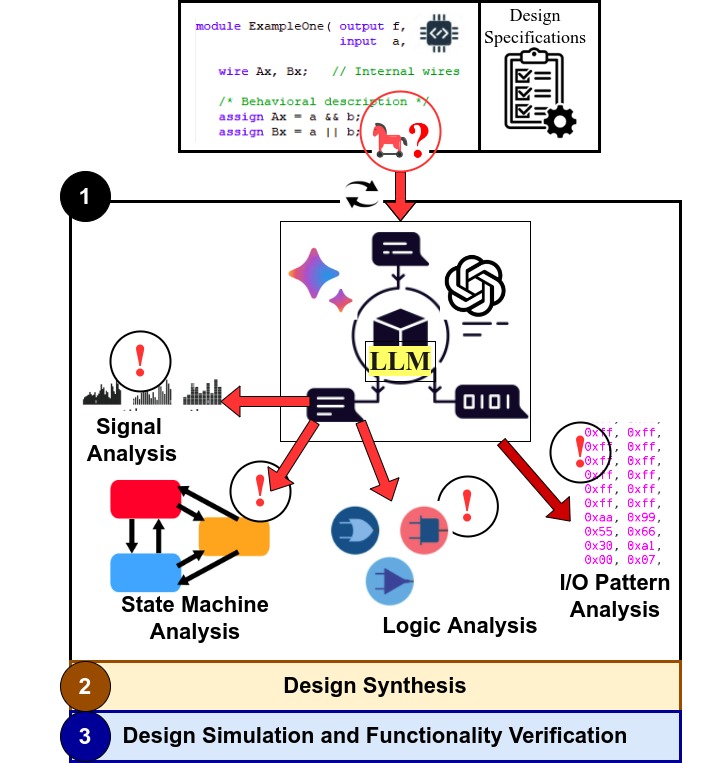}}
\label{fig:attack_model}
\caption{\sentaur{} Flow and its scope in IC Supply Chain.}
\end{figure*}

However, the security and reliability of 3P-EDA/IPs remain uncertain, posing risks when relying on untrusted IPs and EDA tools. This raises concerns about potential malicious alterations within an IC, capable of causing side-channel leakages of sensitive data, functional changes, performance reduction, or denial of service (DoS). These malicious hardware alterations by unauthorized entities within the IC supply chain are commonly known as Hardware Trojans (HTs)~\cite{5406669}, which can even lead to the extraction of secret keys, enabling an adversary to modify the chip’s configurations and gain full control of the chip~\cite{10.1007/978-3-642-33027-8_2}.  The standard design flow of an SoC commencing from design-level specification to physical IP development in a third-party environment is shown in Figure~\ref{fig:attack_model} (a). It highlights various stages within the IC supply chain identified as possible points where HTs could be injected. 
The adaptability of IP cores at higher abstraction levels in the design cycle makes it easy for attackers to insert HTs. Hence, developing detection and countermeasures against HTs at the design phase i.e. at the softIP level is critical~\cite{5604161,9211451}; their elimination becomes costly in later stages. To measure the effectiveness of a countermeasure, designers need the ability to swiftly investigate the attack space susceptible to HT insertion for that design~\cite{5340158}. 

The HT benchmarks on Trust-Hub~\cite{trusthub} are static and limited concerning the diverse nature of SoC development. According to~\cite{10323782}  most Trust-Hub benchmarks rely on unrealistic assumptions. 
They show that $3$-out-of-$83$ HT benchmarks are effective. Further, FPGA/ASIC development process entails designing control and data paths requiring deep hardware expertise. When modifications/additions are frequent to accelerate an algorithm, it can result in frequent obsolescence problems. Unlike software that can be updated or recompiled, IC designs cannot incorporate changes seamlessly. There is a critical need for a platform-independent framework to add functions to an IC design at RTL.

\subsection{Key Contributions}
\sentaur{} is an LLM-based HT assessment flow shown in Figure~\ref{fig:attack_model} (b). Given a design specification and corresponding RTL, we query \sentaur{} to generate HTs based on a HT trigger class, and payload class. Our contributions are:

\begin{itemize}
 \item A novel use of conversational LLMs to generate RTL that is synthesizable using a  HT netlist from Trust-Hub. A recent study ~\cite{10323782} shows that most HTs are not correct and effective after synthesis. We use LLMs to  insert HTs which after synthesis  persist. 
 \item  \sentaur{} is a flexible, versatile, and platform-independent LLM-based toolchain for inserting design templates given functional descriptions of the design.  It can be used by an adversary to insert or analyze the effect of HTs and by a designer to plug in templates.
 \item Validate \sentaur{} flow from an attacker view using the Xilinx platform. We inserted different HTs in the RTL and validated the designs using the Xilinx FPGA. We proposed a mechanism to sanitize and generate HTs from Trust-Hub through \sentaur{} such that functionally correct HTs post-synthesis are generated.  
\end{itemize}

\subsection{Related Works}
In this section, we will discuss the state-of-art research in HT insertion and detection. The very first work in this direction is the development of an extensive repository of HTs available at Trust-Hub~\cite{trusthub}. However, Trust-Hub is restricted to the number of HT circuits that cover only a fraction of the potential landscape for inserting HTs in digital circuits, thereby restricting the development of varied countermeasures. To overcome this limitation TAINT~\cite{8119268} tool is proposed where HT insertion is done at the various stages of the design cycle. However, the tool anticipates that the user will choose the trigger nets based on recommendations provided by the tool itself. In~\cite{8342270} the authors proposed an automated tool  TRIT to insert HTs in a design by configuring various parameters such as the number of trigger nets, the count of rare nets among these triggers, rare-net threshold determined through signal probability of nets, and the selection of payload. Despite expanding the range of inserted HTs, the TRIT methodology cannot identify the best trigger and payload nets. The work~\cite{cryptoeprint:2023/205} proposes a flow that explores Trojans in physical design layouts restricted to ASIC layouts and applicable at the backend stage. Similarly, in~\cite{10103698} Trojan space is explored at the backend stages of FPGA design flow. In~\cite{8306831}, the authors proposed aflow called HAL that inserts countermeasures against Trojan attacks but doesn't explore the Trojan space.

\begin{table}[!htb]
\scriptsize
\centering
\caption{Comparison of Proposed \sentaur{} with State-of-art HT Insertion Tools}
\label{tab:comp}
\begin{tabular}{|c|c|c|c|}
\hline
\textbf{Work} &
  \textbf{Effort} &
  \textbf{\begin{tabular}[c]{@{}c@{}}Trigger Net\\ Selection\end{tabular}} &
  \textbf{\begin{tabular}[c]{@{}c@{}}Learning Time\\ Required\end{tabular}} \\ \hline\hline
TAINT~\cite{8119268}     & Manual    & \begin{tabular}[c]{@{}c@{}}User Selects\\ Trigger Nets\end{tabular}      & NA \\ \hline
TRIT~\cite{8342270}      & Automated & \begin{tabular}[c]{@{}c@{}}Signal Probability\\ of Nets\end{tabular}     & NA \\ \hline
Yu et al.~\cite{9785666} & Automated & \begin{tabular}[c]{@{}c@{}}Transition Probability\\ of Nets\end{tabular} & NA \\ \hline
MIMIC~\cite{cruz2022automatic} &
  Automated &
  \begin{tabular}[c]{@{}c@{}}supervised and\\ generative ML to extract\\ features of Nets\end{tabular} &
  Yes \\ \hline
\begin{tabular}[c]{@{}c@{}}HT \\ Playground~\cite{sarihi2023trojan}\end{tabular} &
  Automated &
  \begin{tabular}[c]{@{}c@{}}RL based location\\ exploration\end{tabular} &
  Yes \\ \hline
\begin{tabular}[c]{@{}c@{}}\sentaur \\
(This Work)\end{tabular}  & 
Automated & 
User Prompts                                                             & No \\ \hline
\end{tabular}
\end{table}

Yu et al.~\cite{9785666} proposed a methodology that identifies rare nets using the transition probability of nets to insert HTs. MIMIC~\cite{cruz2022automatic}  leverages ML to insert HTs. MIMIC ML model was developed by extracting $16$ functional and structural attributes from existing HT samples, creating numerous HTs tailored for specific designs. 
MIMIC process is intricate, involves multiple stages, and requires extensive learning time to train the model. Trojan Playground~\cite {sarihi2023trojan} functions the same way where Reinforcement Learning (RL) is used requiring extensive training of the model. We propose an LLM-based tool flow \sentaur{} that does not involve a time-consuming training process and is user-friendly in generating extensive set of HTs and insert them into an RTL design. A summary of \sentaur{} tool and its benefits are shown in Table~\ref{tab:comp}. 

The rest of the paper is organized as follows: Section~\ref{sec:threatmodel} summarizes the HT threat model, 
~\ref{sec:motive} motivates this study with a use case, followed by
Section~\ref{sec:toolflow1} and Section~\ref{sec:toolflow2}, which details the \sentaur{} flow and capabilities respectively. Section~\ref{sec:4} expounds on the experimental configuration and outcomes, and lastly, Section~\ref{sec:5} presents the conclusions and potential future directions.

\section{\sentaur{}: Tool Flow and Capabilities}\label{sec:3}
\subsection{Hardware Trojan Threat Model}\label{sec:threatmodel}
Our proposed flow \sentaur{} aims to assess malicious possible HT circuits in RTL code, concerning trigger, and payload design. \sentaur{} is capable of analyzing or identifying trigger-based or continuously active HTs in an RTL code given its specification, including those with a payload circuit intended to alter functionality, reduce performance, leak sensitive data, or disrupt service. Our approach focuses on HT insertion during the design stage through scenarios involving deliberate manual manipulation by untrustworthy 3PIP vendors or 3P-EDA tools targeting the RTL netlist.


\subsection{Motivating Use Case}
\label{sec:motive}
Krieg et al. ~\cite{10323782} argue that the assumptions underlying TrustHUB HT benchmarks are unlikely to apply in real-world scenarios. 
In this section, we will show  LLMs can help generate a synthesizable code that reproduces the result as intended. For this, we analyzed the AES-T800 benchmark for our experiment which is also used in~\cite{10323782} to demonstrate that the benchmark violates the correctness and persistence property after synthesizing the netlist. HT insertion in the AES-T800 benchmark  detects a predetermined sequence of input plaintext and then illicitly transmits the AES secret key via a hidden communication channel. 
However, in~\cite{10323782} it has been found that post-synthesis the trigger signal rises, but it appears that it doesn't occur due to the predefined sequence. Instead, it rises right after the circuit reset goes off. Hence using these benchmarks requires manual intervention by hardware experts. They  sanitize the designs to make them synthesizable ensuring usability, correctness, and consistency. 

\begin{figure*}[!ht]
\centering
\centerline{\includegraphics[scale=0.56]{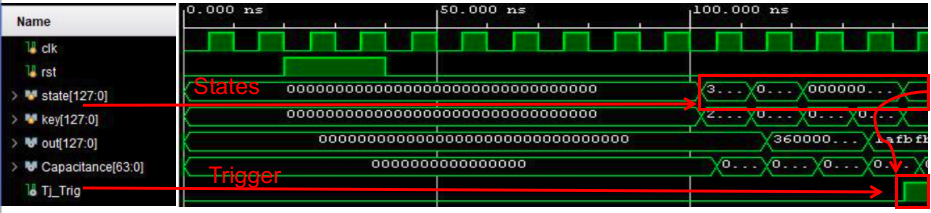}}
\caption{Post-synthesis Simulations of AES-T800 }
\label{fig:llm_post_synth}

\end{figure*}

We leveraged the power of LLM  to generate a synthesizable code efficiently. We chose GPT-4 in our experiment. On prompting GPT-4 as shown below the RTL is generated as shown in Listing~\ref{lst:1254_1}. We synthesized the code using Xilinx Vivado and simulated the post-synthesized using the benchmark from Trust-Hub. The simulated result of the post-synthesized netlist is shown in Figure~\ref{fig:llm_post_synth}; after detecting the sequence of four input states the trigger signal goes high. \\ 

\noindent    \textbf{Prompt:}    \textit{..., once a predefined sequence of $4$ input plaintext ($state$) is observed, alert by setting a signal ($Tj\_Trig$) Write a hardware module in Verilog that is synthesizable.}\\ \\
    \noindent \textbf{LLM Generated Synthesizable Code:}
   \begin{lstlisting}[escapechar=\,language=verilog, caption={Synthesizable AEST-800 HT Benchmark adhering to Correctness and Persistence},label={lst:1254_1}]
 module sequence_detector(input wire clk, input wire rst,  input wire [127:0] state, output reg Tj_Trig);

localparam [2:0] IDLE = 3'b000, SEQ1 = 3'b001,    SEQ2 = 3'b010, SEQ3 = 3'b011,     SEQ4     = 3'b100, TRIGGER  = 3'b101; // Define states
reg [2:0] current_state, next_state;
always @(*) begin // Comb. logic for state transitions
    case (current_state)
        IDLE: 
            if (state == 128'h3243f6a8_885a308d_313198a2_e0370734) 
                next_state = SEQ1;
            else 
                next_state = IDLE;

        SEQ1: 
            if (state == 128'h00112233_44556677_8899aabb_ccddeeff) 
                next_state = SEQ2;
            else 
                next_state = IDLE;

                ...

        SEQ4: 
            next_state = TRIGGER;

        TRIGGER: 
            next_state = IDLE;
// Default case provided by LLM to not infer latch post-synthesis
        default: 
            next_state = IDLE;
    endcase
end

always @(posedge clk) begin
    if (rst) // Sequential logic for state updating
        current_state <= IDLE;
    else
        current_state <= next_state;
end

// Output logic
always @(current_state) begin
    Tj_Trig = (current_state == TRIGGER) ? 1'b1 : 1'b0;
end
endmodule
 \end{lstlisting}
We further expanded the scope and result of this experiment to develop an LLM-based flow \sentaur{} that creates a diverse set of valid HTs based on the functionality of a particular design at the RTL level adhering to correctness post-synthesis. Our tool can be used as a more generic version where it can offer a versatile framework that facilitates automated alterations of provided RTL code to implement specific and predefined functional modifications. 

%% file: files/prompt.tex
\subsection{Tool Flow}\label{sec:toolflow1}
In the last section, we have seen how LLM could generate an RTL code of the Trust-Hub~\cite{trusthub} Trojan benchmark that after synthesis gives functionally correct and intended results. We verified the result using gate-level simulation against the Trojan benchmark provided in Trust-HUb. In this section, we will formalize the flow and introduce an automation framework \sentaur{} that will generate RTL codes given the specification of the RTL design through an LLM as shown in Figure~\ref{fig:attack_model} (b). Our proposed flow is a three-stage process
\begin{enumerate}
    \item \textit{RTL Generation/Analysis using Specification:} The function description of the design under consideration is given to the LLM to generate variations in functionality concerning logic, state machine, I/O pattern, and signal analyses. These functionalities are generated based on some trigger condition specified to LLM based upon time, input-output patterns, or physical conditions. Given that under the purview of HT insertion the generated functionality varies under different trigger conditions, the final effect can be seen as either DoS, leakage of signal values, or degradation of design performance. From a designer's perspective, this feature can be used to plug in templates in a design, and from a defender's perspective, one can use this tool to insert countermeasures to protect against physical attacks. The LLM is directed to generate behavioral descriptions that are synthesizable ensuring functional correctness and persistency at the later stages of the IC design process. Modifications to the behavioral model also enable the tool to be independent of ASIC and FPGA platforms making it a universal tool. The LLM-based generation also enables the flow to be agnostic to vendor-specific file formats, reducing the need for programming or hardware expertise. 
    \item \textit{Design Synthesis:} In this stage the generated RTL design from stage 1 is synthesized using commercial or open-source synthesis tools to generate a gate-level netlist.   
    \item \textit{Design Verification:} The synthesized netlist is verified using a commercial or open-source simulation tool to ensure the desired functionality and correctness.
 \end{enumerate}   
\subsection{\sentaur{} Tool Capabilities}\label{sec:toolflow2}
We will delve into the details of two capabilities in this section and examine how the tool can be utilized in scenarios involving attackers, defenders, and designers.
\subsubsection{Signal Declarations and Connections}
\sentaur{} possesses the capability to incorporate, modify, or eliminate signals, enabling the introduction of new signals tailored for HT-related functions or the alteration of existing signals to include malicious logic. The tool's manipulation of signal connections can be utilized to establish covert communication pathways between the inserted module and external entities. The tool supports various signal modifications:\\
\textit{Input/Output Signals:} The tool augments module declaration by incorporating input/output signals based on user-defined parameters. It caters to scenarios where signals are omitted.\\
\textit{Join Signals:} These signals unify  signals. By replacing the original signal name with the joint signal name, the tool facilitates adding these combined signals into the design. Route and join signals are convenient pathways for transmitting signals that trigger HTs or instructions activating malicious actions.\\
\textit{Add-on Signals:} specified in the extra signals parameter integrate into the module declaration.
\subsubsection{File Manipulation}
\sentaur{} reads the RTL netlist, applies modifications, and saves the modified RTL netlist to the file. This aligns with an attacker's objective of surreptitiously injecting HT logic without raising suspicion.

\sentaur{} offers valuable functionality for modifying HDL files in designs. It can rename modules, manage signal declarations and connections, and manipulate RTL netlists. Designers can customize FPGA/ASIC designs to meet requirements, integrate subsystems, and streamline design processes by utilizing \sentaur{}. Its features grant attackers the ability to obfuscate and embed malicious HTs within FPGA/ASIC designs. Additionally, \sentaur{} aids assessing vulnerabilities in FPGA/ASIC designs and studying impacts of HTs.

%% file: files/results.tex
\section{Experimental Results}\label{sec:4}
\subsection{Setup}
\begin{figure}[th!]
\centering
\includegraphics[width=1\columnwidth]{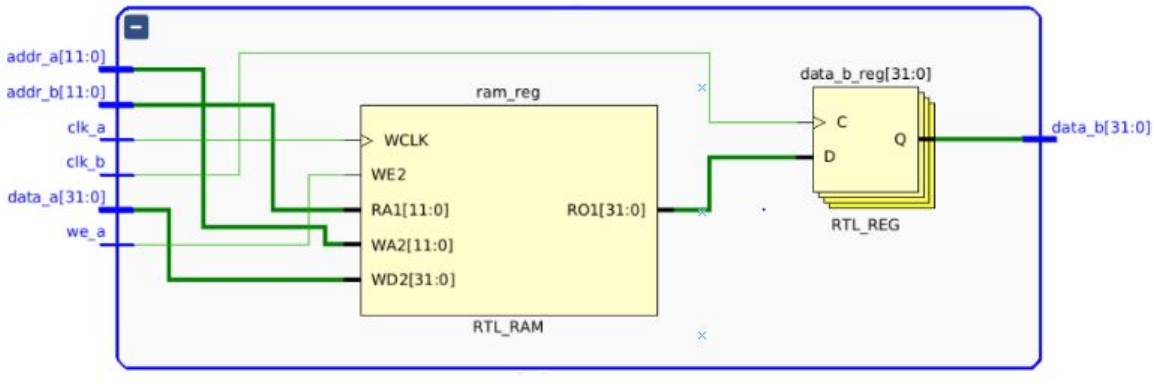}
\caption{Schematic of Dual-port RAM by Xilinx without HT.}\label{fig:without}
\end{figure}

\begin{figure}[th!]
\centering
\includegraphics[width=1\columnwidth]{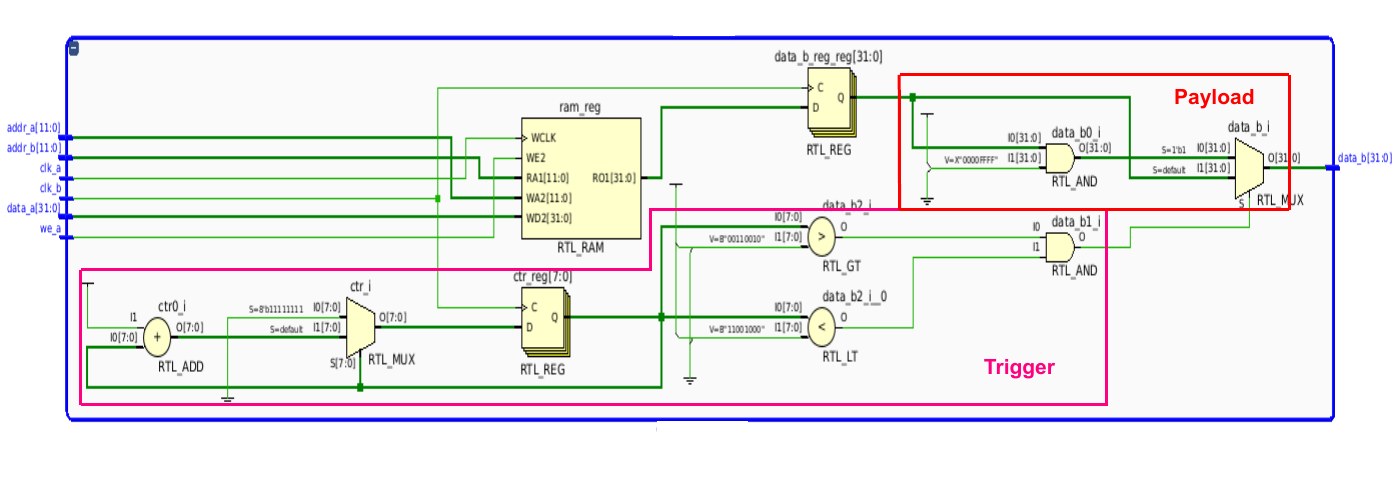}
\caption{Dual-port RAM with HT inserted (State-based) by \sentaur{}. }\label{fig:with}
\end{figure}

For our study, we integrated GPT-4 to our proposed flow \sentaur{} in generating and assessing various HT functionalities. Our RTL benchmark set consists of HTs from Trust-Hub viz. AES-T600, AES-T800, AES-T900, and standard IP dual-port RAM from Xilinx integrated with Processing System (PS) along with additional logic to control the IP. The motivation to choose dual-port RAM for our experiments is that it finds applications in various fields, including digital signal processing, networking, multiprocessing systems, and high-performance computing. Dual-port RAM refers to a type of memory structure that enables two separate read and write operations to occur simultaneously requiring two different clocks. This memory setup allows for independent and simultaneous access to the data stored within it. It's commonly utilized in scenarios where two different processes or modules need access to the same memory resource without causing conflicts or delays.  
For testing the changes done on RTL by our proposed flow \sentaur{}, we employed a real-world Xilinx FPGA of the Zynq-7000 family with device part xc7z020clg400-1. We built a Linux image for the PS with \textit{jupyter notebook} interface compatibility, to control and program the functionality of our SoC design using Python code, with the capability to read and write the memory.
Figure~\ref{fig:without} and Figure~\ref{fig:with} shows the design implemented in the FPGA without and with the HT presence, respectively. This validates that the modification done on top of the RTL by the GPT-4 on the RTL is synthesizable.

\subsection{Result I: Analysis on Generation of HTs}

\begin{figure}[t]
\centering
\includegraphics[width=1\columnwidth]{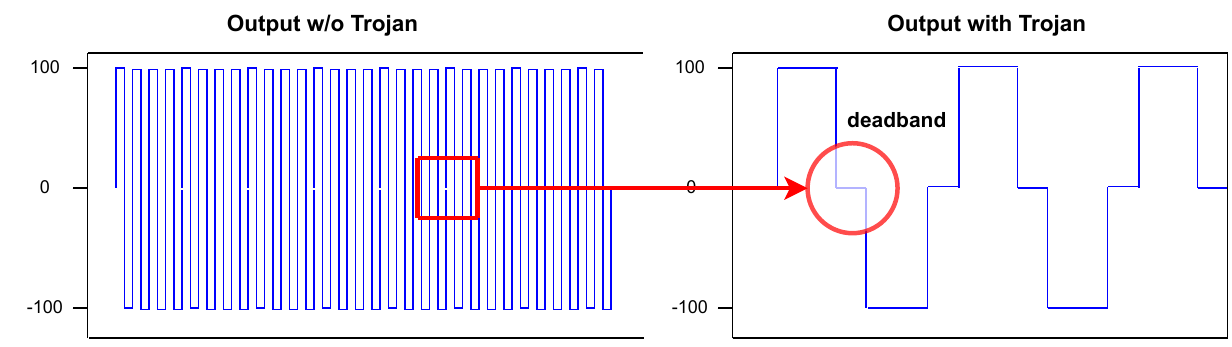}
\caption{Expected outputs with and without  presence (Performance degradation by introducing deadband) when the trigger condition is met.}\label{fig:perf}
\end{figure}

\begin{figure}[t]
\centering
\includegraphics[width=1\columnwidth]{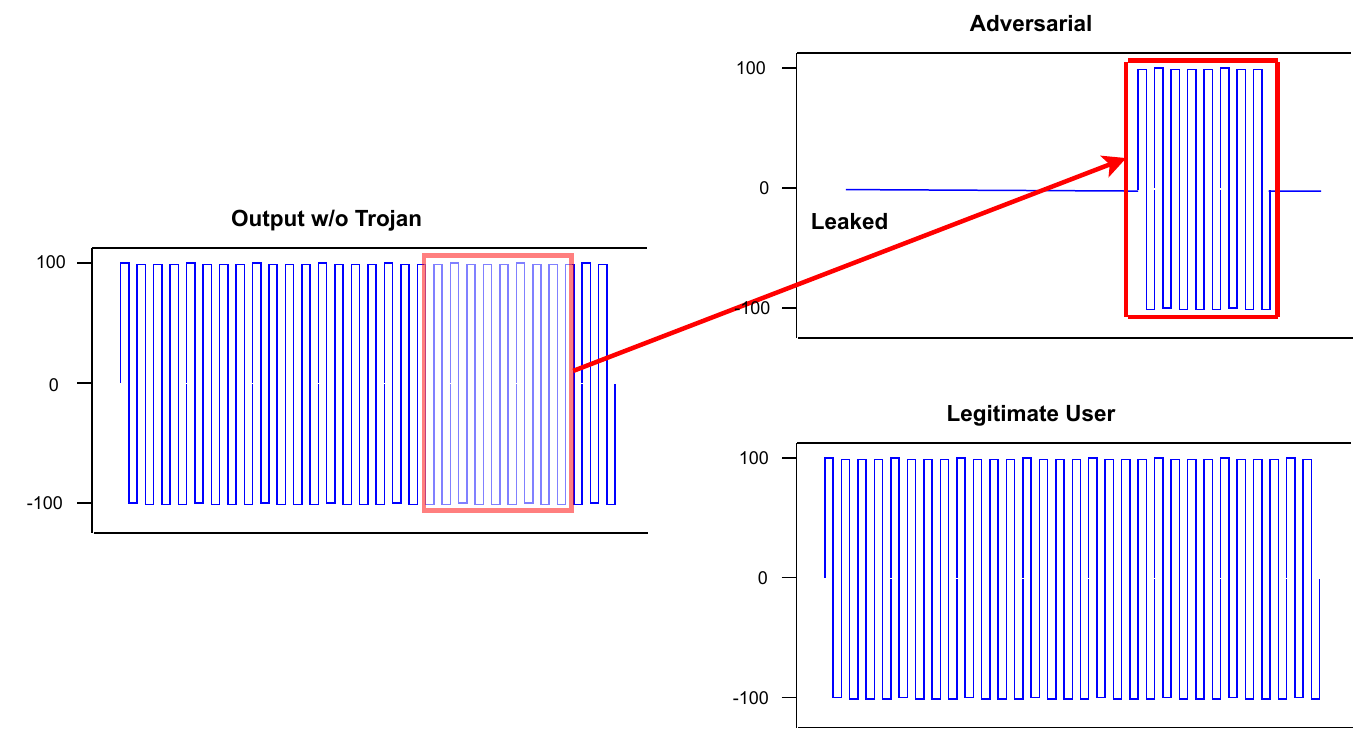}
\caption{Expected outputs with and without  (Information leakage by copying the data to Adversarial) when the trigger condition is met.}\label{fig:leak}
\end{figure}

\begin{figure}[t]
\centering
\includegraphics[width=1\columnwidth]{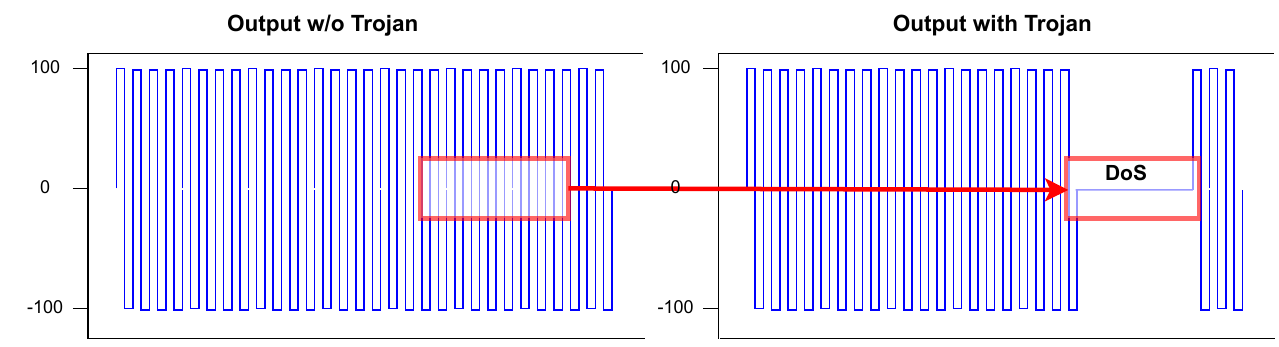}
\caption{Expected outputs with and without  presence (Denial of Service by making the output 0) when the trigger condition is met.}\label{fig:dos}
\end{figure}

\begin{table*}[!ht]
    \centering
    \caption{Result on the Overhead associated with \{using GPT-4 and Manual\}  and the assessment using SENTAUR framework. \\NOTE: $\bullet$ signifies raised concern by the LLM and $\circ$ signifies not flagged by the LLM, respectively. }
    \label{tab:res_gen}

\scriptsize
\setlength{\tabcolsep}{4.2pt}
\begin{tabular}{|c|c|c|c|c|c|c|c|c|c|c|}
\hline
\multirow{2}{*}{Trigger} & \multirow{2}{*}{Description} & \multirow{2}{*}{Effect} & \multicolumn{2}{c|}{GPT-4} & \multicolumn{2}{c|}{Manual} & \multicolumn{4}{c|}{Assessment} \\ \cline{4-11}
& & & LUT & FF & LUT & FF & I/O & FSM & Logic & Signal \\
\hline
\multirow{3}{*}{ Time-based } & \multirow{3}{*}{ When the count reaches in between 50 and 200. } & DOS & 791 & 1261 & 777 & 1080 & $\circ$ & $\circ$ & $\bullet$ & $\bullet$ \\
& & Perf. Degrade. & 846 & 1261 & 829 & 1080 & $\circ$ & $\circ$ & $\bullet$ & $\bullet$ \\
& & Inf. Leak & 798 & 1261 & 779 & 1080 & $\circ$ & $\circ$ & $\bullet$ & $\bullet$  \\
\hline

\multirow{2}{*}{ Logic-based } & \multirow{3}{*}{ When the value of the data is between 1000 and 2000. } & DOS & 772 & 1072 & 774 & 1072 & $\circ$ & $\circ$ & $\bullet$ & $\bullet$ \\
& & Inf. Leak & 778 & 1072 & 778 & 1072 & $\circ$ &$\circ$ & $\bullet$ & $\bullet$ \\
\hline

\multirow{3}{*}{ Address } & \multirow{3}{*}{ When the input address is in the range of 2000 and 3000. } & DOS & 776 & 1072 & 774 & 1072 & $\circ$ & $\circ$ & $\bullet$ & $\bullet$\\
& & Perf. Degrade. & 829 & 1072 & 825 & 1072 & $\circ$ & $\circ$ & $\bullet$ & $\bullet$ \\
& & Inf. Leak & 781 & 1072 & 779 & 1072 & $\circ$ & $\circ$ & $\bullet$ & $\bullet$\\
\hline

\multirow{3}{*}{ State based } & \multirow{3}{*}{ When the FSM detects the sequence 0x55, 0xAA and 0xFF. } & DOS & 785 & 1258 & 821 & 1076 & $\circ$ & $\bullet$ & $\bullet$ & $\bullet$ \\
& & Perf. Degrade. & 837 & 1258 & 853 & 1076 & $\circ$ & $\bullet$ & $\bullet$ & $\bullet$ \\
& & Inf. Leak & 793 & 1258 & 834 & 1076 & $\circ$ & $\bullet$ & $\bullet$ & $\bullet$ \\
\hline

\multirow{3}{*}{ Input based } & \multirow{3}{*}{ When the \# of writes done reaches 1000. } & DOS & 782 & 1104 & 776 & 1084 & $\bullet$ & $\circ$ & $\bullet$ & $\bullet$ \\
& & Perf. Degrade. & 834 & 1104 & 823 & 1084 & $\bullet$ & $\circ$ & $\bullet$ & $\bullet$ \\
& & Inf. Leak & 791 & 1104 & 781 & 1084 & $\bullet$ & $\circ$ & $\bullet$ & $\bullet$ \\
\hline

  AES-T600~\cite{trusthub} &  Detects a specific input plaintext  &  Inf. Leak  &  4588  &  4475  & 4682  &  4521  & $\circ$  & $\circ$  &  $\bullet$  & $\bullet$  \\

\hline

 AES-T800~\cite{trusthub} & Checks for a predefined sequence of input plaintext is observed  & Inf. Leak &  4591  &  4483  &  4726 &  4525  &  $\circ$  &  $\bullet$  & $\bullet$  &  $\bullet$  \\
\hline

AES-T900~\cite{trusthub}  &  After each 2$^{128}$ encryptions, the  gets activated  &  Inf. Leak  &  4771  &  4502  &  4834  & 4523  & $\bullet$  & $\circ$  & $\bullet$  &  $\bullet$  \\
\hline

\end{tabular}
\end{table*}

For this study, 
we took various combinations of triggers and effects for our analysis. Specifically, we took 5 triggers \{time-based, logic-based, address, state-based, input-based\} and 3 effects \{Denial-of-service (DoS), Performance Degradation (Perf. Degrade), Information Leakage (Inf. Leak)\}, respectively. Table \ref{tab:res_gen} gives the overall taxonomy as well as the description of each trigger. For our effects:
\begin{itemize}
 \item \textit{Perf. Degrade} by introducing a dead band which compromises the performance in terms of output being 0 for some small interval at a regular instant (similar to having some delay in between 2 transmissions) with a slight impact in the amplitude as shown in Figure~\ref{fig:perf}.
    \item \textit{Info. Leak} signifies side-information leakage as shown in Figure~\ref{fig:leak} through the use of different channels than the one used for signal transmission, thus making an adversary aware of the data being transmitted to some other users. This can compromise the privacy of a user.
    \item \textit{DoS} by making the output 0 thus denying any service by not transmitting the required data as shown in Figure~\ref{fig:dos}.

\end{itemize}
 We generated the required changes in the RTL by appropriately prompting the GPT-4 with the information of the desired result. To compare the ones generated by the RTL and how a designer would have written them, we compared the LUTs and FFs required by each of them. Table \ref{tab:res_gen} summarizes our result of the different overheads required in both scenarios demonstrating a comparable result when LLM is used. 
 In addition, we took a few examples from the Trust-Hub~\cite{trusthub} specifically, AES-T\{600,800,900\} and did a similar analysis, where we compared the overhead associated while designed by a human and LLM, respectively.

\subsection{Result II: Assessment of HTs}

For the second set of analyses, we focused on manually written HT codes, using LLM (GPT-4), to evaluate these codes for malicious components or indicators of suspicious activity. To ensure a comprehensive assessment, we employed a variety of code combinations as detailed in Table \ref{tab:res_gen} including examples from Trust-Hub~\cite{trusthub}. A key strength of GPT-4 is its ability to summarize and analyze code, allowing it to identify potential vulnerabilities within the code. Crucial to this process is the use of carefully crafted prompts, which guide the LLMs in their analysis. We structured the prompts around four specific areas: logic, state machines, I/O pins, and signal analysis, aligning with the methodology depicted in Figure \ref{fig:attack_model}(b). This targeted approach enabled the harnessing of GPT-4 analytical capabilities to detect and flag areas in the HT codes that might pose security risks.

\begin{itemize}
    \item \textit{I/O pin:} In this analysis, it looks for any I/O pins that are used for some sort of condition (or trigger) that can modify the result in the code.
    \item \textit{State Machine:} In this, the analysis seeks to flag FSMs that look for a particular sequence and depending on that satisfy some condition that can affect the result. There can be many FSMs in a real-world RTL code, which from a human point of view become quite difficult to keep track of, so if LLM can point to only those FSMs where there is some suspicious behavior, it will be of tremendous help from a designer point of view.
    \item \textit{Logic:} This is the most common flag detected by LLMs in various scenarios, indicating distinct logic in the code or dependency on specific conditions to activate. While this might trigger false alarms, focusing on thorough code scrutiny, especially concerning security, outweighs the potential risks of overlooking critical issues.
    \item \textit{Signal:} For this, we look for any potentially vulnerable signal that can trigger some part of the inactive logic in normal operations. LLMs can do a great job in reporting only those signals where a human has to go through the whole code and follow the transition of the signal which becomes cumbersome in a large codebase.
\end{itemize}

Table~\ref{tab:res_gen} summarizes our findings under the column name `Assessment'. It shows the part of the code being flagged as potentially vulnerable by the LLM.

\section{Conclusion and Discussion}\label{sec:5}

\sentaur{} is a framework leveraging an LLM to generate a diverse set of legitimate HTs at the RTL. Unlike existing tools that require a learning period to replicate threat models, \sentaur{} rapidly generates HT instances using LLMs, bypassing the learning phase. It effectively assesses Trust-Hub HTs, conducting comprehensive evaluations with practical use cases and Trust-Hub benchmarks, and exploring diverse impacts and trigger mechanisms at the RTL level. While our primary focus is HT insertion evaluation, \sentaur{} also offers a flexible framework for automated modifications of RTL to integrate specific functionalities. Future direction involves extending this work to insert functionalities post-synthesized netlist.